\documentclass[10pt,aps,prb,twocolumn,amsmath,amssymb,superscriptaddress]{revtex4-1}

\usepackage{graphicx}
\usepackage{amsmath}
\usepackage{amsfonts}
\usepackage{url}
\usepackage{epstopdf}

\begin{document}

\title{Chemistry of the spin-${\scriptstyle\frac{1}{2}}$ kagome Heisenberg antiferromagnet}

\author{Yuan Yao}
\affiliation{Laboratory of Atomic and Solid State Physics, Cornell University, Ithaca, NY 14853-2501, USA}
\author{C. J. Umrigar}
\affiliation{Laboratory of Atomic and Solid State Physics, Cornell University, Ithaca, NY 14853-2501, USA}
\author{Veit Elser}
\affiliation{Laboratory of Atomic and Solid State Physics, Cornell University, Ithaca, NY 14853-2501, USA}

\date{\today}


\begin{abstract}
We believe that a necessary first step in understanding the ground state properties of the spin-${\scriptstyle\frac{1}{2}}$ kagome Heisenberg antiferromagnet is a better understanding of this model's very large number of low energy singlet states. A description of the low energy states that is both accurate and amenable for numerical work may ultimately prove to have greater value than knowing only what these properties are, in particular when these turn on the delicate balance of many small energies. We demonstrate how this program would be implemented using the basis of spin-singlet dimerized states, though other bases that have been proposed may serve the same purpose. The quality of a basis is evaluated by its participation in \textit{all} the low energy singlets, not just the ground state. From an experimental perspective, and again in light of the small energy scales involved, methods that can deliver all the low energy states promise more robust predictions than methods that only refine a fraction of these states.
\end{abstract}
\maketitle

\section{Introduction}

In the past 30 years there has been a surge of interest in the Heisenberg antiferromagnet with spins arranged on corner-sharing triangles in the kagome arrangement. Publications are growing in proportion to their number, with currently over two papers being generated every day. What may have started as idle speculation about the origin of missing entropy in a system of adsorbed He$^3$ atoms and their nuclear spins \cite{greywall1989heat,elser1989nuclear}, the KHA is now a leading candidate for supporting exotic order \cite{sachdev1992kagome,hastings2000dirac}, a driver in the development of numerical methods \cite{singh2007ground,poilblanc2010effective,yan2011spin,lauchli2019s}, and a target for experimental realizations \cite{fukuyama2008nuclear,helton2007spin,jo2012ultracold}. This KHA paper does none of these but instead offers a fresh theoretical perspective along with modest numerical evidence supporting the new approach.

In condensed matter phenomena we are guided by the Landau paradigm, where the low energy physics is derived from general characteristics of the ground state. This strategy, while enormously successful, assumes we have a firm grasp of the ``chemistry" of our system. To see what can go wrong, consider the case of the quantum chemist who undertakes a study of the hydrogen-oxides, in particular, the contentious 2-1 compound. He/she is limited to studies of small clusters, and is frustrated because the ground state properties (structure factor, phonon spectrum, etc.) depend sensitively and unpredictably on system size, boundary conditions, pressure. By going straight to the lowest energy properties of the system, the researcher has failed to notice that the atoms single-mindedly first form H$_2$O molecules, and it is the quirky interactions among these constituents that is responsible for the complex behavior of the bulk compound.

Not meaning to imply a parallel between the KHA and the essential molecules of life, it is at least worth asking whether we have a comparably good understanding of the ``chemistry" of this system of quantum spins. Do we know of a basis of states that provide an accurate representation of the low energy properties, even if a theory for this representation may turn out to be hopelessly complex? After all, there is no comprehensive theory of the 18+ phases of ice other than the physics behind the interactions of water molecules (hydrogen bonds, etc.).

The prevailing strategy for developing a theory of the low energy KHA physics runs counter to the lesson of ice physics. This is the parton (slave-fermion) construction \cite{yoshioka1989slave}, where instead of reducing the entropy of the relevant states it is \textit{doubled}. Notwithstanding the constraint imposed to restore two states per site, this approach is favored because the expanded Hilbert space provides relatively direct access to candidate proposals for ground state order in the mean field approximation. There is also general agreement that these proposals need to be investigated by other techniques, since the reliability of mean field conclusions are questionable when the associated ``large $N$" is only 2 in the original model.

As an alternative to the prevailing strategy we propose the following. First, we shift the focus from divining the KHA's ground state and instead consider its chemistry. The chemistry might turn out to be very interesting, and may even have greater value than establishing ``ice-X" as the ground state. Second, we apply rigorous tests to show that a proposed, reduced-entropy chemical model reproduces the low energy physics. Finally, the computational efficiencies enabled by a validated chemical model give us access to potentially messy questions, including the nature of the ground state. One of the earliest models of the KHA chemistry is featured as an example of the new approach.

\section{Husimi-cactus and spin-singlet dimers}

By not insisting on the perfect kagome topology we can better understand the chemistry of the KHA \cite{elser1993kagome}. The simplest is to arrange the corner-sharing triangles not on the vertices of a honeycomb, but the vertices of an infinite 3-valent tree: the husimi-cactus. Writing the Hamiltonian (in general) as a shifted sum over spins on triangles,
\begin{align}\label{H}
\mathcal{H}&=\sum_\Delta \mathcal{H}_\Delta,\\
\mathcal{H}_\Delta&={\textstyle\frac{1}{2}}\left({\textstyle \sum}_{i\in \Delta}s_i\right)^2-{\textstyle\frac{3}{8}}= \sum_{\langle i j\rangle\in \Delta}s_i\cdot s_j\;+{\textstyle\frac{3}{4}},
\end{align}
we get a zero-energy ground state if we can construct a wave function where each triangle has total spin one-half. There is a two dimensional space of spin-doublets on a triangle, with special linear combinations corresponding to two of the spins forming a singlet, leaving the remaining spin free to form a singlet with a spin on the adjacent corner-sharing triangle.

A completely spin-singlet dimerized husimi-cactus, a ground state of $\mathcal{H}$, is an instance of localization in the following sense \cite{mila1998low}. Whereas there is a one-dimensional continuum of ground states on any one triangle, only a set of three localized settings of that degree of freedom allows the free spin to form a singlet with a spin on an adjacent triangle, and thereby allow this order to propagate through the rest of the cactus.

\begin{figure}[t!]
\center{\scalebox{.4}{\includegraphics{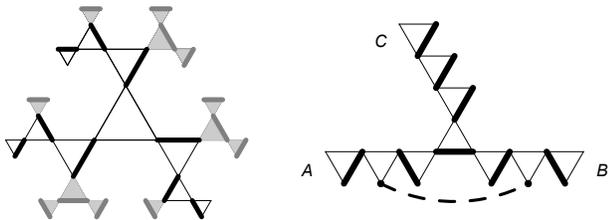}}}
\caption{\textit{Left}: Fragment of the husimi-cactus with a defect triangle (center) in its dimerization. The choice of dimers elsewhere (without defects) determines three semi-infinite chains of (unshaded) triangles on which the Hamiltonian $\mathcal{H}$ acts. \textit{Right}: Interpretation by Hao and Tchernyshyov \cite{hao2009fermionic} of the action of $\mathcal{H}$ as translational motion of two spinons (in a distant singlet relationship) along the same chains (topologically) as in the diagram shown on the left.}
\end{figure}

It is also possible to localize energy on the husimi-cactus \cite{elser1993kagome}. The relevant part of the husimi-cactus is shown in the left panel of Figure 1. One triangle has all three of its spins dimerized with spins on adjacent triangles. This triangle, and the 2-fold choice of dimers along each of the three chains of triangles emanating from it, define a subsystem upon which the action of $\mathcal{H}$ is confined (the singlets on the shaded triangles in Figure 1 remain undisturbed). Because the ``defect triangle" defined by the intersection of the three semi-infinite chains does not have a singlet pair, it fails to be an energy eigenstate. However, starting with the state $\Psi_0$ shown in Figure 1 we can construct (by Lanczos) a sequence of orthogonal basis states $\Psi_1,\Psi_2,\ldots$ generated by successive applications of $\mathcal{H}$, each disrupting the dimerization one step further down the chain. From these we obtain estimates $E_d(0)=0.75$, $E_d(1)=0.5$, $E_d(2)=0.459$, $E_d(3)=0.444$, etc. for the defect triangle energy as we expand the basis. Hao and Tchernyshyov \cite{hao2009fermionic} showed that these converge to $E_d(\infty)=0.378$ and established that the excitation is localized. We should note that not only can this energy be placed on any triangle of the husimi-cactus, but there are exponentially many (in the number of triangles) ways for the three chains to meander through the cactus.

The KHA is usually described as a ``frustrated" system, where the presence of triangles defeats the N\'eel alignment of classical spins. But on the husimi cactus, with the help of the spin-dimer localized basis, we see that this system is not frustrated at all. Though Anderson \cite{anderson1973resonating} long ago proposed a resonating spin-dimer (``valence bond") basis for another classically frustrated system, the triangular lattice, we believe it is in the localized dimer setting that this basis confers an advantage over other bases. As we describe below, the KHA is frustrated in a very different way, and in contrast to the husimi-cactus, by the fact that defect triangles of the kind described above are not excitations but imposed by topology.

The work of Hao and Tchernyshyov (HT) \cite{hao2009fermionic} advanced the chemical understanding of the KHA in an important way. HT interpreted the defect triangle on the cactus as a bound state of two spinons. A spinon on the cactus, where one spin is not dimerized and all other spins form dimers, one per triangle, is another zero energy state. There is no zero energy two-spinon state, but there are positive energy states where two spinons are confined to the same three-pronged set of triangles as the defect triangle in the left of Figure 1. A basis state is shown in the right panel and we see that the dimer environments in which the spinons find themselves are not eigenstates at the triangles on which they reside. The action of $\mathcal{H}$ in this case not only admixes further-neighbor singlets but also generates translations of the spinons along the chain of triangles. When one spinon is restricted to chain $A$, the other spinon is constrained to move along chains $B$ or $C$, etc. Also, when the two spinons exchange position in this manner, HT noticed that the dimer wave function changes sign, conferring fermi statistics to the spinons. The spins of a spinon pair can be combined into a singlet or triplet, and HT find the singlet combination has the lower energy, binding the spinons in close proximity to the defect triangle. The singlet to triplet excitation energy, $\Delta E_1\approx 0.06$ \cite{hao2009fermionic}, is very small and makes spinon unbinding a strong candidate for the unusually small $\Delta E_1$ observed numerically for the KHA \cite{iqbal2014vanishing,he2017signatures,lauchli2019s}.

Although the husimi-cactus has the same local geometry as kagome, the two systems deviate in an important way with respect to a topological property of the spin-dimerized states. For any dimerized state, including states with spinons, there is a rule for assigning a $\pm 1$ flux to all the edges of the ``triangle-graph" upon which the triangles are placed (3-valent tree, honeycomb) \cite{elser1993kagome}. This gives the triangles a net charge, and the low energy ``singlet triangles" ($\mathcal{H}_\Delta=0$)
all have charge -1.
The net flux entering the system, in a low energy state, must therefore grow in proportion to the enclosed number of triangles. This is only possible in graphs, such as trees, where the number of edges crossing the boundary scales with the number of vertices interior to the boundary.

\begin{figure}[t!]
\center{\scalebox{.4}{\includegraphics{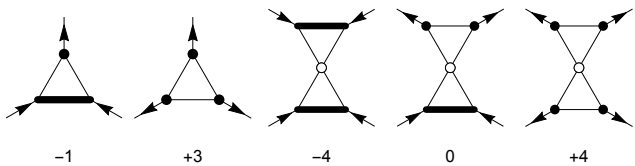}}}
\caption{Fluxes (arrows) and corresponding charge $Q$ of the two kinds of triangle and three environments of a spinon. Black vertices not in a dimer are dimerized with a spin on the adjacent triangle (not shown).}
\end{figure}

The ``arrow rules" for assigning fluxes and corresponding charges $Q$ are shown in Figure 2 for the two kinds of triangle in a fully dimerized state, as well as the three kinds of triangle environments of an isolated spinon. Arrows are associated only with sites that have spins and point toward the triangle that contains the spin's dimer partner. In the case of spinons the charge is assigned to triangle pairs by the net exiting flux. Of the zero energy configurations, the $Q=-1$ single triangle is superior to the $Q=-4$ triangle-pair because it minimizes the accumulation of charge, when the system has nonextensive boundary. As sources of neutralizing countercharge, the contenders are the $Q=+3$ defect triangle and the $Q=0$ spinon environment named the ``anti-kink" by HT \cite{hao2009fermionic}.  The former increases the charge (over the $Q=-1$ background) by $\Delta Q=4$ at energy cost $\Delta E=E_d$, while the anti-kink has $\Delta Q=2$ and energy equal to half the unbound spinon-pair energy, $\Delta E=(E_d+\Delta E_1)/2$. Of these, the defect triangle has the smaller value of $\Delta E/\Delta Q$, by an amount proportional to $\Delta E_1$.

\section{Charge neutral systems and loops}

Systems with nonextensive boundary are topologically frustrated and have positive energy (relative to the husimi-cactus) from the finite density of charge-neutralizing defect triangles. From their $+3$ charge relative to the $-1$ charge of defect-free triangles, we know the defect triangle concentration is fixed at $1/4$. The presence of loops in the triangle network represents another point of departure from the husimi-cactus. In the tree topology, different dimerized states (including ones with defects) are related by infinite chains of triangle edges along which the two states choose a different alternating sequence of dimers. By contrast, in the KHA these chains can be finite loops, making the dimerized states nonorthogonal. As a result, the Hamiltonian now not only ``dresses" the environments of the defect triangles but also mediates transitions in their positions.

For any hexagon in the triangle graph of the KHA, and any dimerized state, there is a unique transition-loop to another dimerized state that encircles only the given hexagon. These transition loops generate all the dimerized states, and from this we know their number is $2^{N_\ell}$, where $N_\ell$ is the number of loops (hexagons). An early proposal \cite{elser1993kagome} for constructing a low energy effective Hamiltonian for the KHA was based on the generalization where the hexagons in the triangle graph are replaced by polygons with $s$ sides. In such a system without boundary, analogous to the KHA with periodic boundary conditions, the (3-valent) triangle graph has $N_\Delta$ vertices, $N=(3/2)N_\Delta$ edges (spins), $N_\ell=(2/s)N$ polygons, and lies in a surface of genus
\begin{equation}
g=\left(\frac{s}{6}-1\right)\frac{N_\ell}{2}+1
\end{equation}
by Euler's theorem. The same rules for assigning charges to vertices and fluxes to the edges of the triangle graph apply in this generalization, including the concentration of defect triangles, $N_d=N_\Delta/4$. For $s=7$ the smallest system has 12 loops, 28 triangles and 42 spins.

\begin{figure}[t!]
\center{\scalebox{.45}{\includegraphics{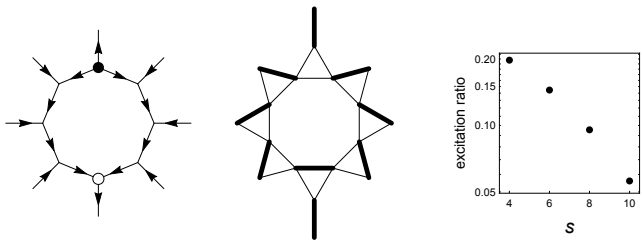}}}
\caption{Resonance on $s$-gons, shown for $s=8$, is generated by the reversal of arrows around the ring (left panel) and moves a defect from the top triangle to the bottom triangle in the dimerization (center panel). The ratio of the two lowest singlet excitation energies, $\Delta E_{1\to 2}/\Delta E_{1\to 3}$, is plotted on the right as a function of $s$.}
\label{resonance}
\end{figure}

The $s$-gon generalization of the KHA clarifies its relationship to the husimi-cactus model and disentangles the diagonal and off-diagonal terms for the $N_\ell$ pseudo-spin variables $\boldsymbol\sigma$ in the effective Hamiltonian. For the state with $\sigma^z=+1$ on all the polygons we may pick any valid dimerization/arrow-assignment. Flipping a pseudo-spin corresponds to reversing the arrows on just its polygon, as shown in Figure \ref{resonance} for $s=8$. The flux out of a polygon varies from only inward arrows to any even number of outward arrows. In the former case there are two defect-free dimerizations with exact local energy degeneracy. Otherwise, alternating out-arrows give the locations of defect triangles, switching roles in the two states. Resonance now splits the energies of the two dimerization by an amount we expect to scale as their overlap, $(1/2)^{s-d}$, where $d\le s/2$ is the number of defects. To show that we  recover a two-level system in the limit of large $s$, we numerically obtained the two lowest singlet excitation energies for the case $d=1$, where we Heisenberg-coupled the spins at the $s$-gon's out-arrows to a pair of spins in the polygon's environment. The ratio $\Delta E_{1\to 2}/\Delta E_{1\to 3}$, shown in Figure \ref{resonance}, decays exponentially with $s$ and is already quite small for $s=6$. In absolute terms, the resonance energy gain of the two lowest singlets, $T=\Delta E_{1\to 2}/2=0.029$, is also small for $s=6$.

Whereas resonance splitting disappears for large $s$, the $2^{N_\ell}$ pseudo-spin states continue to acquire different energies through the positions of the defect triangles. Repeating the Lanczos defect triangle calculation on the husimi-cactus, now for a pair, we find the energy is lowest when the pair is at their closest separation (one intervening triangle), but only by about $V=-0.01$. This is consistent with the high order dimerized-coupling perturbation theory calculation of Singh and Huse \cite{singh2007ground}.

In addition to learning that both diagonal and off-diagonal terms of the effective Hamiltonian $\mathcal{H}_\mathrm{eff}$ are small, the exercise of looking at the model for general $s$ has shown us that the form of the Hamiltonian is complicated. Using $V_\ell\{\sigma^z\}$ to denote a general function on the set of $z$-pseudo-spins on the loops (polygons) adjacent to loop $\ell$, on which we have pseudo-spin $\boldsymbol\sigma_\ell$, the effective Hamiltonian takes the following form,
\begin{equation}\label{Heff}
\mathcal{H}_\mathrm{eff}=\sum_\ell\left( {\sigma^z}_\ell\;V_\ell\{\sigma^z\}+{\sigma^x}_\ell\;T_\ell\{\sigma^z\}\right)+\cdots\;,
\end{equation}
where the omitted terms are higher order in the number of flipped pseudo-spins.
The first term is able to count the number of nearest defect triangle pairs, each contributing $V\approx-0.01$. This is because every nearest defect triangle pair has an associated loop $\ell$, and ${\sigma^z}_\ell$ along with the $\sigma^z$ of the adjacent loops specify the existence and positions of all the defect triangles on $\ell$. Likewise, the resonance energies $T_\ell$ are a function of the number and positions of the defect triangles around the loop, which are specified by the adjacent $\sigma^z$. For example, when there is a single resonating defect and $s=6$, $T_\ell= 0.029$.

The $s$-gon generalization of the KHA has helped us identify relevant small energies in its chemistry. Shifting $\mathcal{H}$ as in \eqref{H} to make the ground state energy lower-bounded by zero (the husimi-cactus energy), the excess energy per triangle of the KHA is only $\Delta E_0/N_\Delta=0.0919$ \cite{lauchli2019s}. This number can be compared to the energy of a $1/4$ concentration of defect triangles, $E_d/4=0.0945$, and is consistent with the observation that defect-defect interactions ($V$) and resonance gains ($T$) are both small. That $\Delta E_0/N_\Delta$ itself is a small number should remind us that the KHA is only weakly ``frustrated" in the basis of spin dimerized states. Increasing $s$ quickly reduces $T$, and the ground state selects dimer configurations that minimize just the diagonal terms, $V$, which remain unchanged and small. The generalization of the KHA for large $s$ is also interesting insofar as spinons are out of the picture. Though resonance may be interpreted as the unbinding of spinon pairs at all the defect triangles around an $s$-gon, and their subsequent recombination at the intervening out-arrow positions, the low energy states are well described without any reference to spinons.
Finally, whereas $s\to \infty$ is formally the husimi-cactus, on which the dimerized states are truly localized, we should not expect this to be the case for any finite $s$. Dynamics/thermalization will be slow at large $s$ (even $s=6$), but not frozen.

The program to analyze the KHA via the Hamiltonian $\mathcal{H}_\mathrm{eff}$ was abandoned when a numerical study \cite{zeng1995quantum} revealed that including higher order resonance terms brought a qualitative change to the ground state properties. Apparently $s=6$ is not sufficiently large for $\mathcal{H}_\mathrm{eff}$ to capture all details of the chemistry relevant for the ground state. At the lowest level of resonance a translational symmetry-broken state is weakly favored, in fact the same state identified by the high order Heisenberg-coupling perturbation analysis of Singh and Huse \cite{singh2007ground}.

\section{Testing the spin-dimerized basis}\label{SDB}

To make progress on the KHA we propose setting aside, for now, the resolution of the ground state and the derivation of an effective Hamiltonian, and instead shift the focus to the low energy ``chemistry." Numerical studies \textit{do} agree on two things: (\textit{i}) the triplet gap $\Delta E_1$ is small, but nonzero, and (\textit{ii}) there are unusually many low energy singlet states. Building on these findings might proceed as follows. First, we define ``low energy singlet" as any state with energy below $\Delta E_1$. Clearly, being able to determine the number of low energy singlets, $N_0$, for a given small system, would demonstrate our command of the chemistry. Also, it is hard to imagine how that goal can be achieved without at the same time having the capability to construct good bases for the low energy singlets. We evaluate a basis $B$, in its representation of a low energy state $\Psi$, by its \textit{participation} $p=|P_B(\Psi)|^2$, where $P_B$ is the projection to the span of $B$. In systems small enough where it is feasible to numerically obtain all the low energy singlets $\Psi$, a single number that quantifies its quality is the participation averaged over all $\Psi$, denoted $\overline{p}$.

We illustrate the new approach with the basis of dimerized states, $D^0$. As in two earlier studies \cite{elser1993kagome,zeng1995quantum}, we refine this basis by admixing further-neighbor singlets generated by $\mathcal{H}$ at each defect triangle. That is, for each near-neighbor dimerized state $\Psi$ we construct the (unnormalized) basis state
\begin{equation}\label{D0}
\Psi(\alpha)=\prod_\Delta(1+\alpha \mathcal{H}_\Delta)\Psi,
\end{equation}
where $\alpha$ is a variational parameter. Fluctuations generated by a single application of $\mathcal{H}$ are responsible for most of the energy reduction of an undressed defect triangle on the husimi-cactus. When there are multiple defects, fluctuations to this order are  independent of the defect configuration because defects can never be on adjacent triangles.

\begin{figure}[t!]
\center{\scalebox{.4}{\includegraphics{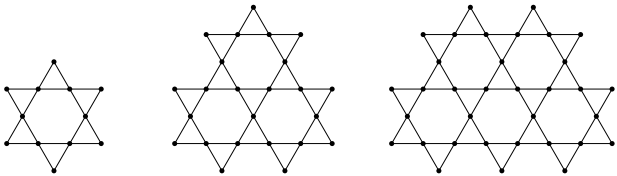}}}
\caption{First three bounded clusters with an even number of spins, comprising 1, 3 and 5 hexagons.}
\label{clusters}
\end{figure}

To test the quality of our basis we use the family of bounded clusters shown in Figure \ref{clusters}, where the triangle graph is comprised of the union of an odd number of hexagons. From a chemical perspective, clusters with boundary are better for testing the versatility of a basis and avoid the artifact of short, nonzero winding number transition-loops, when small systems are placed on a torus. Not being focused on the ground state of the infinite system is another reason periodic systems hold less sway.

Starting with the cluster built on one hexagon with 12 spins, it is easy to see that attaching another hexagon always adds an odd number of spins. In this study we are interested in the low energy singlet states and therefore keep the number of hexagons (elementary loops) $N_\ell$ odd. To study the dynamics of a single spinon one would use systems with even $N_\ell$. Since transition-loops around the hexagons uniquely generate all the basis states, our basis has size $2^{N_\ell}$. Finally, by taking into account that the triangles on the boundary with only two corner-sharing neighbors always have charge $Q=0$ in a dimerized state, it is easy to work out, from charge neutrality, that the number of defect triangles $N_d$ satisfies $N_d=(N_\ell-1)/2$.

Table 1 summarizes our results for the odd $N_\ell$ clusters up to the $N_\ell=5$ cluster with 34 sites. For the two larger clusters we used a custom-parallelized version of the Lanczos program \cite{spectra_library}  to find energy eigenstates up to at least the first spin triplet, with the energies converged to below $1\times10^{-10}$. The dimer bases $D^0$ were optimized with respect to $\alpha$, \textit{not} to get the best ground state energy, but to maximize the average participation $\overline{p}$ of the singlets below the lowest triplet. Shown also are results for the augmented bases $D^1$ of size $2\times 2^{N_\ell}$ obtained by including states generated by a single application of $\mathcal{H}$ to $D^0$. We see that the values of $\overline{p}$ remain large even for the largest system.

\begin{table*}[htb]
\caption{Summary of results for the three systems in Figure \ref{clusters} using the complete basis and the
dimer bases $D^0$ and $D^1$.
$N_\ell$ is the number of hexagon loops, $N_d$ the number of defects, $\Delta E_0/N_\Delta$ the excess energy
per triangle (over the husimi-cactus), $\Delta E_1$ the singlet-triplet gap, $N_0$ the number of singlet states below the lowest
triplet, $\alpha$ is a variational parameter, and $\overline{p}$ the average participation.}
\begin{tabular}{|cc|ccc|ccc|ccc|}
\hline
\multicolumn{2}{|c|}{}&\multicolumn{3}{c|}{complete basis}&\multicolumn{3}{c|}{$D^0$}&\multicolumn{3}{c|}{$D^1$}\\
\hline
$N_\ell$ & $N_d$ & $\Delta E_0/N_\Delta$ & $\Delta E_1$ & $N_0$ & $\alpha$ & $\overline{p}$ & $\Delta E_0/N_\Delta$ & $\alpha$ & $\overline{p}$ & $\Delta E_0/N_\Delta$ \\
\hline
1 & 0 & 0 & 0.259669 & 2 & --- & 100.\% & 0 & --- & 100.\% & 0\\
3 & 1 & 0.028009 & 0.132053 & 7 & -0.3961 & 90.9\% & 0.034501 & -0.2745 & 95.2\% & 0.031055\\
5 & 2 & 0.039642 & 0.098374 & 13 & -0.4071 & 83.5\% & 0.046416 & -0.2815 & 90.2\% & 0.042458\\
\hline
\end{tabular}
\end{table*}

\begin{figure}[htb]
\center{\scalebox{.45}{\includegraphics{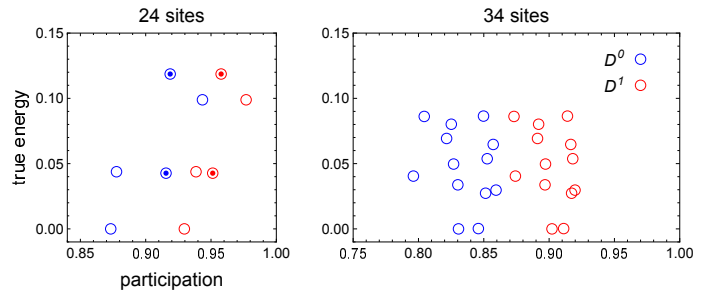}}}
\caption{Participation $p$ of all the singlets below the lowest triplet state in the bases $D^0$ and $D^1$. The bullseye symbols represent pairs of degenerate states allowed by the symmetry of the $N_\ell=3$ cluster.}
\label{partplot}
\end{figure}

\begin{figure}[t!]
\center{\scalebox{.45}{\includegraphics{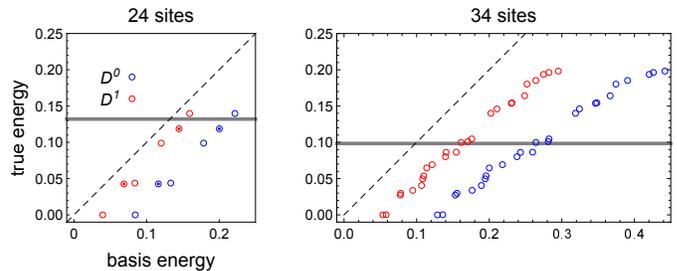}}}
\caption{Energies of the lowest singlet states computed in the complete basis compared against their values in the bases $D^0$ and $D^1$. The gray horizontal line shows the energy of the lowest triplet state.}
\label{spectra}
\end{figure}

Evidence that the quality of our bases applies uniformly to all the low energy singlets is shown in Figure \ref{partplot}. The effect of going from the basis $D^0$ to the doubled basis $D^1$ is a nearly uniform shift to higher participation $p$. A basis with high participation will also give an accurate spectrum when the Hamiltonian is projected onto it. This is shown in Figure \ref{spectra}, where points on the dashed diagonal correspond to perfect representation by the basis. When the scatter of points is nearly parallel to the diagonal, it means the basis is doing a good job representing the density of states.

\section{Alternative bases}

To our knowledge, two other bases for the low energy states have been proposed. Like our bases, these too were inspired by high degeneracy ground states for particular modifications of the KHA model. Mila \cite{mila1998low,mambrini2000rvb} considered partitioning the Heisenberg couplings into sets of strength $J$ and $J'$, such that $J'\to 0$ results in just the kagome ``up-triangles" being internally coupled and decoupled from each other. Perturbation theory for $J'\ll J$ is complicated, motivating Mila to consider the basis of singlet states obtained by forming singlet dimers of the spin-doublets on adjacent ``trimers" of kagome spins. The size of the resulting basis, corresponding to dimers on the triangular lattice formed by the kagome up-triangles, grows as $1.154^N$, where $N$ is the number of kagome spins.

Preserving translational symmetry, in contrast to Mila, Changlani and co-workers \cite{changlani2018macroscopically,changlani2019resonating} modified the KHA by moving from the Heisenberg point $J_z/J_\perp=1$ to the special anisotropic case  $J_z/J_\perp=-1/2$. The basis is now given by all tensor products of three spin states in the familiar 120$^\circ$ relationship, with the constraint that adjacent kagome sites have different  spin states, or ``colors." The degeneracy of the zero $z$-magnetization sector is believed to be the same as the number of kagome 3-colorings, $1.134^N$, and it is this sector (after projection to total spin zero) that is of interest for the Heisenberg model.

The trimerization and 3-coloring bases offer a clear advantage in economy over our husimi-cactus-inspired basis, which grows as $2^{N/3}=1.260^N$. However, the former bases break symmetries of the KHA and it is harder to make the case they have the precision required for the small energies in the model. In general, the average participation $\overline{p}$ of an incomplete basis decays exponentially with the number of spins. It would be useful to know how symmetry breaking compromises bases in this  respect, relative to a basis that does not. Systems with boundary (Fig. \ref{clusters}) also present challenges. At boundaries, down-triangle spins not also part of an up-triangle are left isolated in a trimerization. The 12-site system, with its two dimer basis states (both exact ground states), has 11 permutation-inequivalent 3-colorings.

\section{Evidence of a low energy sector}

The small energy scales of the KHA make its low energy properties exceptionally sensitive to realities beyond the model (disorder, anisotropy, etc.), thereby complicating efforts to test ground state hypotheses through experiment. A more robust experimental indicator of kagome physics would be evidence of an unusually high concentration of states at low energy.
The ``missing entropy" question that was raised 30 years ago \cite{greywall1989heat} did in fact receive a satisfactory resolution in the He$^3$ system \cite{fukuyama2008nuclear}, with heat capacity measurements at lower temperatures. However, Roger \cite{roger1990frustration} pointed out that more elaborate spin models, that naturally arise in ring-exchange systems like solid He$^3$, could also explain the observed double-peaked heat capacity.

Without making assumptions about the nature of the ground state order, Elstner and Young \cite{elstner1994spin} convincingly showed there was indeed low energy structure in the KHA heat capacity by combining a high temperature series with the spectra of small systems up to 18 spins. The current state-of-the-art along these lines is the study by Schnack and co-workers \cite{schnack2018magnetism} that finds evidence, using the finite temperature Lanczos method on systems up to 42 sites, of a broad heat capacity ``shoulder" that extends to two orders of magnitude below the temperature of the main peak. While there is still much to be resolved experimentally and numerically for the KHA, the general phenomenon of an abundance of low energy states deserves a theoretical explication. The development of high quality bases for the low energy states seems to us as the first logical step in that direction.

\section{Computation with low energy bases}

After control over the quality of a low energy basis is demonstrated, through its participation in a sufficiently large set of energy eigenstates in the full basis, subsequent calculations can take advantage of the economies provided by the greatly reduced basis size. However, size reductions generally incur extra costs, such as nonorthogonality of the spin-dimerized bases. These extra costs and their growth are assessed in the appendix for the block-Lanczos method. The latter is a general technique that can exploit the property of an initial basis being already reasonably good, so that far fewer iterations are needed than in conventional Lanczos with random initial states. The doubling of the basis in Sec. \ref{SDB}, from $D^0$ to $D^1$,  is an example of a single block-Lanczos iteration.

The chief downside in the application of block-Lanczos to low energy bases is the rapid growth in memory with iterations. This is mitigated by the very small memory requirement for the initial basis. We illustrate this point with the example of our spin-dimerized bases. Consider a KHA system on the torus comprising $N$ spins, $N_\ell=N/3$ hexagon loops and $N_d=N_\ell/2$ defect triangles. The initial basis has $M=O(2^{N_\ell})$ states, each of which requires memory for $4^{N_d}=2^{N_\ell}$ elementary spin-dimerized states by the 4-fold multiplication at each defect triangle in \eqref{D0}. Thankfully, each elementary spin-dimerized state uses only $O(N)$ memory to store (symbolically) the matching of the spin pairs. Overall, the initial basis therefore requires $O(M^2)=O(2^{2N/3})$ memory, an exponential improvement over the complete basis.

The blocks in the block-Lanczos method are $M\times M$ matrices of numbers that represent $\mathcal{H}$ in the block-tri-diagonalized form. Storage of these blocks does not pose a problem. What does pose a problem is the growth in the number of (symbolic) elementary spin-dimerized states with each application of $\mathcal{H}$. When the Lanczos blocks are kept dense, the multiplication factor in each iteration is $M$, cancelling the memory savings over the full basis after just one iteration. Fortunately, the spin-dimerized basis has the nice feature that the Lanczos blocks are naturally near-sparse. Matrix elements decay exponentially in the number of flipped pseudo-spins by which the corresponding basis states differ. By limiting this number, though more generously than the single flipped pseudo-spin of the early attempt at an effective Hamiltonian \cite{elser1993kagome}, the memory growth can be managed. This can be implemented by setting an absolute threshold that the block matrix elements must exceed in magnitude to be retained.

\section*{Acknowledgements}

We thank Hitesh Changlani, Debanjan Chowdhury and David Huse for helpful discussions on the content, David Mermin for improving the presentation, and dedicate this work to the memory of Chris Henley.
This work was supported in part by the AFOSR under grant FA9550-18-1-0095 and by
the Molecular Sciences Software Institute under U.S. National Science Foundation grant ACI-1547580.
The computations were performed under Google Cloud Platform's Research Credits program and at the Bridges cluster at the Pittsburgh Supercomputing Center supported by NSF grant ACI-1445606.

\section*{Appendix: Block-Lanczos for the spin dimer basis}

This appendix introduces notation and supports claims made in the main text about the complexity of implementing the block-Lanczos algorithm for the spin-dimer basis. Block-Lanczos, when used with a special initial block, can take advantage of the ``chemistry" of the system and deliver good results with far fewer iterations than when used on an arbitrary initial basis. We make an effort to distinguish those parts of the implementation that are specific to the spin-dimer basis from those that apply to bases more generally.

Spin-dimerized states, from the perspective of computation, are best understood as symbolic objects. An elementary spin-dimerized state $|\phi\rangle$, for a system with an even number of spins $N$, is completely specified by a matching $\delta$ of the integers $\{1,\ldots, N\}$, that is, a map on this set with the properties $\delta(i)\ne i$, $\delta^2(i)=i$ :
\begin{equation}\label{elementary}
|\phi\rangle=\prod_{i=1}^N\frac{\mathrm{sgn}(\delta(i)-i)}{2^{1/4}}\;\left(\,|+\rangle_i\,|-\rangle_{\delta(i)}\;-\;|-\rangle_i\,|+\rangle_{\delta(i)}\,\right).
\end{equation}
Here $|+\rangle_i$ denotes an up-spin at site $i$, etc. To store this state in the symbolic sense we only need memory for the $N$ integers $\delta(1),\ldots, \delta(N)$. A general spin-dimerized state, given by a sum of $K$ elementary states
\begin{equation}
|\psi\rangle =\sum_{k=1}^K\alpha_k\,|\phi_k\rangle,
\end{equation}
requires memory for $K N$ integers and $K$ complex numbers.

The symbolic representation of spin-dimerized states is sufficient for both of the operations we need to perform: acting with the Hamiltonian $\mathcal{H}$ and computing inner products. Inner products distribute over the elementary states and for these
\begin{equation}
\langle\phi_1|\phi_2\rangle=(-1)^{r(\delta_1,\delta_2)}\;2^{c(\delta_1,\delta_2)},
\end{equation}
where the integers $r$ and $c$ are easily computed from the associated matchings $\delta_1$ and $\delta_2$. The action of $\mathcal{H}$ on the elementary dimerized states is also very simple and in fact reminds us why this basis was chosen in the first place. Consider the term $\mathcal{H}_\Delta$ in $\mathcal{H}$, where the triangle $\Delta$ comprises spins $(i,j,k)$ :
\begin{equation}
\mathcal{H}_\Delta=\frac{1}{2}(P_{i j}+P_{j k}+P_{k i}).
\end{equation}
Here $P_{i j}$ exchanges the site labels $i$ and $j$ in the product \eqref{elementary}, etc. Now if $\delta(i)=j$, or $\delta(j)=k$, or $\delta(k)=i$ in the state $|\phi\rangle$, then $\mathcal{H}_\Delta |\phi\rangle=0$, and indeed the action of this part of $\mathcal{H}$ is very simple. If none of these apply, then in the case of the first term we may assume $\delta(i)=m\ne j$ and $\delta(j)=n\ne i$, where $m\ne n$. The action of $P_{i j}$ on $|\phi\rangle$ is the exchanges $\delta(i)\leftrightarrow\delta(j)$ and $\delta(m)\leftrightarrow\delta(n)$ in the matching $\delta$, possibly with a sign change applied to the amplitude of the state. In the worst case, when no triangles have a dimer in the state $|\phi\rangle$, $\mathcal{H}|\phi\rangle$ will be a sum of $2N$ elementary dimerized states, where $2N$ is just the number of Heisenberg couplings (exchange operators) in a system of $N$ spins (assuming a system without boundary). The number $2N$ is therefore the worst case growth factor, for each application of $\mathcal{H}$, in the memory requirement for general dimerized states.

For a system with no boundary on the torus and $N_\ell$ loops in the triangle graph, the initial basis has $M=2^{N_\ell-1}$ spin-dimerized states in each topological sector. With a slight abuse of notation we define
\begin{equation}
\widetilde{B}_0=[\;|\psi_1\rangle\; \cdots \;|\psi_M\rangle\;]
\end{equation}
as the rectangular matrix of basis vectors, the $M$ ``columns" of which are understood as being symbolic in their representation. In our basis $D^0$ for the KHA, each $|\psi_i\rangle$ is the result of applying the factor $(1+\alpha\,\mathcal{H}_\Delta)$ to each of the $N_d$ defect triangles of a single elementary dimerized-state. The memory requirement for each column of $\widetilde{B}_0$ is therefore $O(4^{N_d})=O(2^{N_\ell})=O(M)$, since $N_\ell=2 N_d$ and the memory for an elementary dimerized state is sub-exponential in $N$. The Cholesky decomposition,
\begin{equation}
(t_0)^\dag t_0 =(\widetilde{B}_0)^\dag \widetilde{B}_0,
\end{equation}
of the $M\times M$ matrix of inner products $\langle\psi_i|\psi_j\rangle$, defines an upper triangular matrix $t_0$ with which we can construct an orthonormal basis by
\begin{equation}\label{B0}
B_0=\widetilde{B}_0\, (t_0)^{-1}.
\end{equation}

The first block-Lanczos iteration is defined by the equation
\begin{equation}\label{L1}
\mathcal{H}B_0 = B_0\, h_0 +B_1\, t_1,
\end{equation}
where the lower-case $M\times M$ matrices $h_0$ and $t_0$ should be seen as forming linear combinations of the columns of the bases $B_0$ and $B_1$, while $\mathcal{H}$ on the left side acts symbolically on the columns of $B_0$. Basis $B_1$ is uniquely defined up to phases when we insist it is orthonormal and orthogonal to $B_0$. Applying these  properties to \eqref{L1} we obtain
\begin{subequations}\label{iter0}
\begin{eqnarray}
h_0&=&(B_0)^\dag(\mathcal{H}B_0)\\
(t_1)^\dag t_1&=&(\mathcal{H}B_0)^\dag (\mathcal{H}B_0)-(h_0)^2\\
B_1&=&((\mathcal{H}B_0)-B_0\, h_0)(t_1)^{-1},
\end{eqnarray}
\end{subequations}
where $t_1$ is upper-triangular, analogous to $t_0$. The general block-Lanczos iteration $i=1,2,\ldots\;$ is defined by
\begin{equation}\label{Li}
\mathcal{H}B_i = B_{i-1}\,(t_i)^\dag + B_i\, h_i +B_{i+1}\, t_{i+1},
\end{equation}
where the first term on the right is implied by the hermiticity of $\mathcal{H}$. Analogous to \eqref{iter0} we now have the following three steps in the iteration:
\begin{subequations}
\begin{eqnarray}
\!\!h_i&=&(B_i)^\dag(\mathcal{H}B_i)\\
\!\!(t_{i+1})^\dag t_{i+1}&=&(\mathcal{H}B_i)^\dag (\mathcal{H}B_i)-(h_i)^2-t_i(t_i)^\dag\\
B_{i+1}&=&((\mathcal{H}B_i)-B_i\, h_i-B_{i-1}\,(t_i)^\dag)(t_{i+1})^{-1}. \nonumber \\
\end{eqnarray}
\end{subequations}

We have already commented on the fact that the term $\mathcal{H}B_i$ in the block-Lanczos recursion has a hidden complexity growth coming from the multiplication of the number of symbolic terms in the columns of $B_i$ when acted upon by $\mathcal{H}$. Another, and  more serious growth in complexity is associated with the terms where a basis is right-multiplied by a numerical matrix, such as $h_i (t_{i+1})^{-1}$. The latter is a dense $M\times M$ matrix and will in the worst case multiply the number of symbolic terms in each column of $B_i$ by $M$. This growth is much more rapid than the growth caused by the action of $\mathcal{H}$, and will impose a severe limit on the number of iterations unless mitigating measures are taken.

We can use the prior knowledge that the matrices $h_i$ and $t_i$ have a hierarchy of magnitudes to make the block-Lanczos algorithm practical. First consider the starting (nonorthogonal) basis $\widetilde{B}_0$. The dimerizations in any two of its columns differ by arrow-reversals on some number of the hexagons in the triangle graph. In the language of the effective Hamiltonian for pseudo-spins \eqref{Heff}, two columns of $\widetilde{B}_0$ differ by some number of flipped values of $\sigma^z$. The off-diagonal elements of $(\widetilde{B}_0)^\dag \widetilde{B}_0$ accordingly decay exponentially with the number of flipped pseudo-spins between the two states/columns. Because an earlier study \cite{zeng1995quantum} showed that truncating the off-diagonal elements at just one flipped pseudo-spin was too severe, we should consider a parameterized truncation scheme that admits off-diagonal elements for multiple flipped pseudo-spins. The simplest such scheme is to impose an absolute threshold $\epsilon$ on the magnitude of the retained elements of the Cholesky factor $t_0$ and its inverse, thereby controlling its sparsity. This limits the growth in the size of the columns of the orthonormal basis \eqref{B0}. The same threshold principle to control sparsity can be applied to $h_0$ and, in the general recursion, $t_i$ and $h_i$.

In other bases (trimerization \cite{mila1998low}, 3-coloring \cite{changlani2018macroscopically}), whose sizes $M$ are smaller than the dimerized basis for the same system size, the growth in the sizes of the basis states with Lanczos iteration will accordingly be less of a problem. Even so, memory growth by a factor $M$ per iteration quickly becomes impractical and thresholding to impose sparsity is a necessity. We note that the spectra for the spin-dimerized basis $D^1$ shown in Figure \ref{spectra} correspond to no threshold ($\epsilon=\infty$) and $k=1$ Lanczos iterations, that is, where the Hamiltonian for the low energy singlets is represented by a block-tri-diagonalized matrix with dense blocks $h_0$, $h_1$ and $t_1$. Because our interest there was basis participation in the full basis, sparsity considerations were not relevant.
In larger systems, when states cannot be refined in the full basis, a low energy basis can still be assessed with respect to its convergence to an unknown spectrum. The computational cost for this convergence will then depend both on the sparsity threshold $\epsilon$ of the matrix blocks, as well as the number of Lanczos iterations $k$.

\bibliographystyle{apsrev}
\bibliography{kagome}

\end{document}